\def\b{B_{\rm tur}}
\def\B{B_{\rm reg}}
\def\disk{^{\rm (D)}}

\def\DPT{{\rm DP}_{20.5/2.8}}

\def\dz{\Delta z}
\def\err#1#2{{{{\hfill#1}\atop{\hfill#2}}}}
\def\ers#1{\,{\scriptstyle{\pm#1}}}
\def\H{B}
\def\halo{^{\rm (H)}}
\def\hsyn{h_{\rm syn}}
\def\hth{h_{\rm th}}
\def\HI{{\rm H\,{\scriptstyle  I}}}
\def\HII{{\rm H\,{\scriptstyle  II}}}

\def\nelectron{n_{\rm e}}
\def\R{{\cal R}}
\def\RM{{\rm RM}}

\def\RMfg{\rm RM_{\rm fg}}

\def\sun{_{\odot}}
\def\cm{\,{\rm cm}}
\def\cmcube{\,{\rm cm^{-3}}}

\def\mJyb{\,{\rm mJy/beam}}
\def\K{\,{\rm K}}
\def\kpc{\,{\rm kpc}}

\def\mkG{\,\mu{\rm G}}

\def\p{\,{\rm pc}}
\def\radm{\,{\rm rad\,m^{-2}}}

\def\sun{_\odot}

\def\half{{\textstyle{1\over2}}}

\def\AA{{\rm A\&A}}
\def\ApJ{{\rm ApJ}}
\def\ARAA{{\rm ARA\&A}}
\def\MNRAS{{\rm MNRAS}}
\def\Nat{{\rm Nat}}
\def\PASJ{{\rm PASJ}}
\def\heidel{Beck R., Kronberg P.P., Wiele\-binski R. (eds.)
     Proc. IAU Symp. 140, Galactic
     and Intergalactic Magnetic Fields. Kluwer, Dordrecht}
\def\leiden{Bloemen H. (ed.) Proc. IAU Symp. 144, The Interstellar
    Disk-Halo Connection in Galaxies. Kluwer, Dordrecht}
\def\potsdam{Krause F., R\"adler K.-H., R\"udiger G. (eds.) Proc. IAU
    Symp. 157, The Cosmic Dynamo. Kluwer, Dordrecht}
\def\pheins{\phantom{1}}
\def\phbrack{\phantom{(}}

  \MAINTITLE={Magnetic fields in the disk and halo of M51}
  \SUBTITLE={ ????? }
  \AUTHOR={ E.M.\ts Berkhuijsen@1,
            C.\ts Horellou@2,
            M.\ts Krause@1,
            N.\ts Neininger@1
            A.D.\ts Poezd@3,
            A.\ts Shukurov@4
       \newline and
            D.D.\ts Sokoloff\ts@3
         }
    \OFFPRINTS={ E.M.\ts Berkhuijsen }
  \INSTITUTE={
@1 Max-Planck Institut f\"ur Radioastronomie, Auf dem H\"ugel
        69, 53121 Bonn, Germany
@2 Onsala Space Observatory, S--43992 Onsala, Sweden
@3 Physics Department, Moscow University, Moscow 119899, Russia
@4 Computing Center, Moscow University, Moscow 119899, Russia }
\DATE={Received March 4, 1996, accepted June 1996 }
%
\ABSTRACT={
We discuss the {\it regular\/} magnetic field in the galaxy M51
on the basis of a new interpretation of polarization angles observed
at the wavelengths $\lambda\lambda$2.8, 6.2, 18.0 and 20.5\ts cm.
We found a {\it magneto-ionic halo\/} in M51  with a radial extent
of about 10\ts kpc. The regular magnetic fields in the disk and the
halo have different structures. The regular magnetic field in the
halo is axisymmetric and horizontal. Its field lines are spirals
pointing {\it inwards\/} and generally opposite to those in the disk.

The azimuthal structure of the magnetic field in the {\it disk\/} is
neither axisymmetric nor bisymmetric but can be represented by a
superposition of these two basic harmonics with about equal weights.
Magnetic lines of the regular field in the disk are spirals generally
directed {\it outwards\/}. They are well aligned with the optical
spiral arms as traced by the dust lanes.

The regular magnetic field strength averaged in 3\ts kpc wide rings
is about 5--10$\mkG$ in the disk and $3\mkG$ in the radial range
3--6\ts kpc in the halo. These values are in good agreement with
independent estimates from the synchrotron emission. The general
features of the magnetic patterns revealed in the disk and the halo
seem to be in agreement with predictions of dynamo theory.

Our results are also based on an analysis of the magneto-ionic medium
in M51. For a range of radii in the galactic disk, we estimated the
scale heights of synchrotron and thermal disks, electron densities
and filling factors as well as the amount of depolarization.
}
\KEYWORDS={spiral galaxies -- interstellar magnetic field -- radio
        continuum -- polarized radio emission -- Faraday rotation --
        M51 -- depolarization }
\THESAURUS={ 03(09.13.1; 11.19.2; 11.09.1 M51, 13.18.1; 03.20.5)  }
\maketitle
\MAINTITLERUNNINGHEAD{Magnetic fields in the disk and halo of M51}
\AUTHORRUNNINGHEAD{E.M.\ts Berkhuijsen et al.}
\titlea{Introduction}

\begfigwid 18.0 cm\
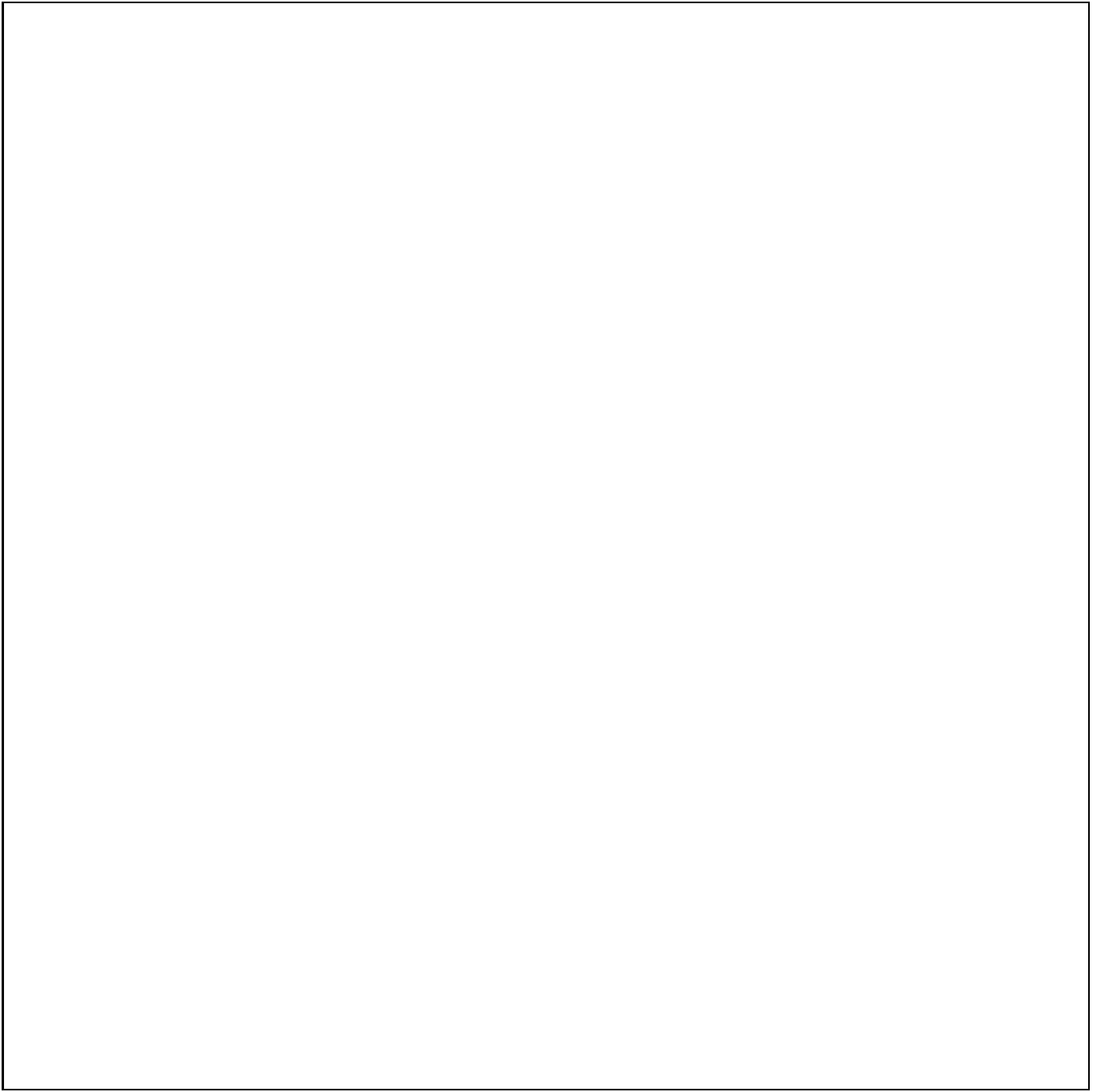
\figure{1}{ Maps of the E-vectors rotated by 90\degr\  observed at
$\lambda\lambda$2.8\ts cm (a), 6.2\ts cm (b), 18.0\ts cm (c) and
20.5\ts cm (d). The length of the vectors is proportional to the
observed polarized intensity. They are shown superimposed onto an
optical picture (Lick Observatory) of M51. All maps were smoothed to
an angular resolution of 75\arcsec }
\endfig

M51 (NGC~5194) is the first external spiral galaxy from which
linearly polarized radio emission was detected (at $\lambda20$\ts cm
by Mathewson et al.\ 1972, and at $\lambda\lambda6.0$ and 21.2\ts cm
by Segalovitz et al.\ 1976), and for which the global structure of
the regular magnetic field highlighted by these observations was
investigated (Tosa \& Fujimoto 1978). It is also the first galaxy
for which the discovery of the so-called bisymmetric magnetic
structure was claimed (Tosa \& Fujimoto 1978), a configuration of
magnetic field which was later suspected for some other nearby spiral
galaxies (see Krause 1990, Beck 1993 and Beck et al.\ 1996 for a
review). The explanation of the origin of the bisymmetric magnetic
fields has become one of the major challenges to the theory of
galactic magnetic fields of the last decade.

Later observations of M51 with better sensitivity and resolution
(Neininger 1992a; Horellou et al.\ 1992; Neininger et al.\ 1993a;
Neininger \& Horellou 1996), and also a more careful analysis of the
observational data, revealed a magnetic pattern which is considerably
more complicated than a simple bisymmetric structure. This is true
also for some other galaxies, e.g. M83 (Neininger et al.\ 1991,
1993b).

One of the goals of the present paper is to introduce and test a
general way to represent the complicated magnetic patterns observed
in galaxies in terms of a tractably small number of parameters.
Hopefully, this will facilitate a fruitful confrontation of theory
with observations. Such a {\it parametrization\/} can be conveniently
performed in terms of the Fourier expansion of the magnetic field in
azimuthal angle. The lowest Fourier harmonic corresponds to
the axisymmetric magnetic field, the next higher one to the
bisymmetric mode, etc. However, we emphasize that the Fourier
harmonics thus derived are not necessarily connected with the dynamo
modes and their physical meaning should be established using models
of the magnetic field generation and evolution in a given galaxy.

We attempted to build a coherent, self-consistent picture of the
global magnetic structure based not only on the Faraday rotation
analysis, but also on other available information coming from, e.g.,
intrinsic polarization angles, depolarization data, total synchrotron
emission, thermal radio emission, the morphology of the galaxy, etc.
In order to interpret a polarization pattern in a galaxy, one should
know certain parameters of the interstellar medium such as scale
heights of the thermal and synchrotron disks, electron volume density
and filling factor. These are discussed in Sect.~3.

On pursuing our goals we used recent multifrequency observations of
M51, which allowed us to distinguish two magneto-ionic layers along
the line of sight (the disk and halo) with significantly different
magnetic fields.

We adopted the following parameters of M51: centre coordinates $\rm
\alpha_{50} = 13^h27^m46\fs 327,\  \delta_{50} = +47\degr 27\arcmin
10\farcs 25$ (Ford et al.\ 1985), a position angle of the major axis
of $-10\degr$ measured counterclockwise from north, an inclination
angle $i=-20\degr$ ($i=0\degr$ is face-on -- see also Appendix~A in
the electronic version) (Tully 1974), and a distance to M51 of 9.7\ts
Mpc (Sandage \& Tammann 1974).

In this text equations, figures and tables are numbered
as in the electronic version, whereas the sections are numbered
consecutively.

\titlea {The observational database}

The observations of the spiral galaxy M51 which we discuss below were
obtained at the wavelengths 2.8\ts cm (Neininger 1992a), 6.2\ts cm
(Neininger et al.\  1993a), 18.0\ts cm and 20.5\ts cm (Horellou
et al.\ 1992). In contrast to earlier discussions of these observations,
we analyze here the observations for all four wavelengths {\it
simultaneously\/}.

The 2.8\ts cm data are single-dish measurements obtained with the
100-m Effelsberg radio telescope. The other data sets were obtained
with the VLA in its D-array. This imposes some restrictions for the
use of the 6.2\ts cm data since at this wavelength the diameter of
the primary beam of the VLA is 9~arcmin, which corresponds to a
diameter of 25\ts kpc in the plane of M51.  Therefore, we considered
the measurements at 6.2\ts cm to be reliable up to a radius of 9\ts
kpc chosen to be somewhat smaller than the radius of the primary
beam, and we did not use them at larger radii.

At all wavelengths the data were smoothed to a final resolution of
$75\arcsec$ corresponding to $3.5\times3.8\kpc$ in the plane of M51.
Fig.~1 shows the observed E-vectors rotated by 90\degr , superimposed
onto an optical picture of M51.

Following a usual procedure in the studies of regular magnetic fields
in external galaxies, the galaxy was divided into several rings and we
considered the values of polarization angle averaged in sectors in
each ring, $\psi_{ni}$, and the corresponding uncertainties,
$\sigma_{ni}$, where the subscript $n$ refers to the
wavelength and $i$ to the sector. Here we chose the rings between the
galactocentric distances $r = 3$, 6, 9, 12 and $15\kpc$; within each
ring sectors of an opening angle of 20\degr\ were used. The
azimuthal angle $\theta$ was measured counterclockwise from the
northern major axis. Throughout the paper we specify sectors by their
median value of $\theta$. A detailed discussion of the data averaged
in sectors and their uncertainties is given in Sect.~2.2 of the
electronic version.

\begtabfullwid
\tabcap{1}{The synchrotron disk in M51}
\halign{\hfill #\hfill\qquad &\hfill#\hfill\qquad &\hfill#\hfill\qquad
        &\hfill#\hfill\qquad &#\hfill\qquad
        &#\hfill\qquad &\hfill #\hfill\qquad &\hfill#\hfill\cr
\noalign{\hrule\medskip}
$r$ &$I_{2.8}$ &$P_{2.8}$ &$B$ &$\B$  &$\b$ &$h_{2.8}$ &$h_{20.5}$\cr
[kpc] &[mJy/beam] &\omit &[$\mu$G] &[$\mu$G] &[$\mu$G] &[pc] &[pc]\cr
\noalign{\medskip\hrule\medskip}
\pheins 3--\pheins 6 &10.9
 &0.09 &21.1       &6.2 &20.2            &\pheins500 &\phantom{2}800\cr
\pheins 6--\pheins 9 &\pheins 5.9
 &0.18 &15.8       &6.8 &14.3            &\pheins700 &1100\cr
\pheins 9--12 &\pheins 2.6
 &0.12 &12.7       &4.5 &11.9            &\pheins900 &1600 \cr
12--15    &\pheins 1.1
 &0.14 &\pheins9.2 &3.5 &\phantom{2}8.6  &1300       &2200 \cr
\noalign{\medskip\hrule} }
\endtab

\begtabfullwid
\tabcap{2}{Thermal disk and halo in M51}
\halign{#\hfill\qquad\quad
        &#\hfill$\pm$&#\hfill\qquad\quad
        &\hfill# \qquad
        &#\hfill$\pm$&#\hfill\qquad\quad
        &\hfill# \qquad
        &\hfill#  \cr
\noalign{\hrule\medskip}
       &\omit&\omit
                      &\multispan3 \hfill Disk  \hfill
                      &\multispan2 \hfill Halo  \hfill \cr
      &\omit&\omit
                &\multispan3 \hskip-0.5em\hrulefill  \qquad
                & \multispan2 \hskip-0.5em\hrulefill \cr
\hfill$r$   &\multispan 2\hskip 0.8truecm $S_{2.8}$\hfill
            &$\hth$\hfill
            &\multispan 2\hskip 2.5em $\langle\nelectron\rangle$\hfill
            &$Z$\hfill   &$\langle\nelectron\rangle$\hfill
                                     \cr
\hfill [kpc]  &\multispan 2 [mJy/beam]\hfill
             &[pc] \hfill &\multispan 2\hskip 1.5em [$\!\cmcube$]\hfill
               &[pc] \hfill &[$\!\cmcube$] \hfill  \cr
\noalign{\medskip\hrule\medskip}
\pheins 3--\pheins 6 &\phbrack$2.34$&$1.15$
                   &$400$ &\hskip 0.7em $0.11$&$0.03$
                          &$5000$   &$0.003$    \cr
\noalign{\smallskip}
\pheins 6--\pheins 9 &\phbrack$0.98$&$0.35$
                   &$600$ &\hskip 0.7em $0.06$&$0.01$
                          &$3300$   &$0.003$      \cr
\noalign{\smallskip}
\pheins 9--12&\phbrack$0.11$&$0.60$
                   &$1400$ &\hskip 0.7em $0.013$&$0.010$
                          &0        &0            \cr
\noalign{\smallskip}
12--15&$(0.012$&$0.60)$
                   &$2000$
                   &\multispan 2 $(0.004\err{+0.008}{-0.003})$\hfill
                          &$0$   &$0$           \cr
\noalign{\medskip\hrule\medskip}
\noalign{
\parshape=1 3pt 370pt  
\noindent
{\petit Note:\
the negative error in $S_{2.8}$
was cut off to prevent $\langle\nelectron\rangle < 0$.}
\hfill}
\noalign{\medskip\hrule}
}
\endtab


\titlea {The magneto-ionic medium in M51}
\titleb{ The nonthermal disk in M51 }
The exponential scale heights of the synchrotron emission at $\lambda
2.8\cm$ and  $\lambda 20.5 \cm$ given in Table~1, $h_{2.8}$ and $h_{20.5}$
respectively, were estimated from those in the Milky Way by scaling the
latter values obtained at 408\ts MHz (Beuermann et al.\ 1985) with
frequency as $\nu^{-0.25}$, as observed for NGC~891 (Hummel et al.\ 1991a)
and M31 (Berkhuijsen et al.\ 1991). As M51 and the Milky Way are galaxies
of a similar type (Sc and Sbc, respectively) and have about the same linear
dimensions, the scale height in the Solar neighborhood ($r\sun=8.5\kpc$)
was assumed to apply at $r=9\kpc$ in M51.

The total field strength $B$ (including the regular and turbulent
components) can be evaluated from the nonthermal radio intensity per beam
area at $\lambda2.8\cm$, $I_{2.8}$, using the standard
assumption of equipartition of energy density between the magnetic field
and the cosmic rays (see, e.g., Krause et al.\ 1984). The nonthermal
emission was obtained from the total emission by subtracting the thermal
component derived by Klein et al.\ (1984).

Using the total nonthermal intensity one first estimates the strength
of the transverse magnetic field (i.e. the projection of $\vec B$
on the plane of the sky). From the polarized intensity, or the degree of
polarization, one obtains the strength of the transverse regular
magnetic field ${\B}_\perp$, from which one gets $\B$ by deprojection
assuming that the field lies in the galaxy's plane.  The turbulent
magnetic field strength $\b$ is found from its transverse component
by multiplication by $\sqrt3$ assuming statistical isotropy. As
Faraday effects are negligible at $\lambda$2.8\ts cm, the degree of
polarization at this wavelength, $P_{2.8}$, yields the best estimate
for the strength of the regular magnetic field. Therefore we
evaluated $B$, $\B$ and $\b$ from the $\lambda$2.8\ts cm data as
given in Table~1.

We stress that the estimates of $\B$ and $\b$ from the synchrotron
emission are used only to assess the role of Faraday
depolarization effects in Sect.~3.3.  The analysis of the
polarization pattern performed in Sect.~4 yields {\it independent\/}
estimates of $\B$ which are in agreement with those given in Table~1.

\titleb{The thermal disk in M51}

Klein et al.\ (1984) derived the radial dependence of the thermal
radio emission in M51 at $\lambda2.8\cm$. For each ring the average
thermal flux density $S_{2.8}$ per beam area is given in Table~2. To
calculate the average electron density we used the formulae for the thermal
flux density (in terms of emission measure) of $\HII$ regions of Mezger \&
Henderson (1967) in the form given by Israel et al.\ (1973) for an
unresolved source. In our case we have
$$
{{\langle\nelectron\rangle}\over{\cmcube}}
        = C f^{1/2}\left({{S_{2.8}}\over{\mJyb}}\right)^{1/2}
            \left( {{2\hth}\over{\p}}\right)^{-1/2}\;,
\eqno(1)
$$
where $\langle\nelectron\rangle$ is the average electron density in
the thermal ionized gas layer of exponential scale height $\hth$,
$f$ is the volume filling factor of the electron density defined by
$\langle\nelectron\rangle^2=f \langle\nelectron^2\rangle$, and $C$ is
a certain constant depending on distance, resolution and electron
temperature. For a distance of 9.7~Mpc, a resolution of $75\arcsec$,
and a temperature of $10^4\K$ we have $C=7.4$.

Using the values of $f$ and $\hth$ derived below we found the average
electron densities presented in Table~2.

The scale height $\hth$ is not known for M51. Again we can make an
estimate using the values known for the Milky Way. Observations
indicate that $\hth$ varies with radius.

For the inner Galactic region we used the observation of Reich \&
Reich (1988) that the full halfwidth of the thermal emission at
$r<7$\ts kpc is 2\fdg 0--2\fdg 5, which yielded a scale height of
about 400\ts pc at a mean radius of $r=4\kpc$. This scale height was
adopted for the ring 3--6\ts kpc in M51.

For the solar neighbourhood Reynolds (1991a) suggested a
two-component model of the thermal disk. For our present purpose it
is sufficient to consider a one-component model of the thermal disk
described by a single exponential fitted to the sum of the thin and
the thick disk (see Sect.~3.2.2 of the electronic version of the
paper). Then the scale height in the Solar neighborhood, scaled to
$r_{\sun} =8.5\kpc$, is $\hth = 600\p$; this value was taken
for 6--9\ts kpc in M51.

\begfigside 8.0cm 12cm
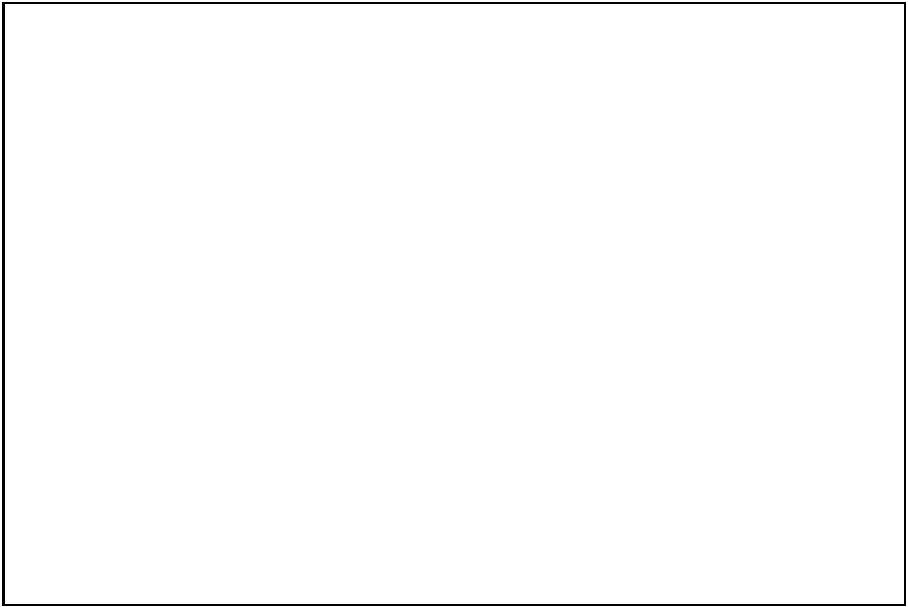
\figure{2}{ Sketch of the various layers in M51 seen in a vertical
plane through the centre: disk of thermal electrons with scale
height $\hth$, synchrotron disk at $\lambda20.5\cm$ with scale height
$h_{20.5}$, and the elliptical halo. The depth in the thermal disk
$\dz$ from which polarized emission at $\lambda$20.5\ts cm is
observed, is indicated. $Z$ is the extent of the halo at $z > \hth$.
For clarity the thermal and synchrotron disk are shown thicker than
to scale.}
\endfig

For the outer regions of M51 we scaled the values derived from
a comparison of $\HII$ and $\HI$ observations of the Milky Way.
The $\HI$ layer in the Milky Way becomes thicker at large radii,
and the same may hold for the ionized gas. Dickey \& Lockman (1990)
derived a constant $\HI$ scale height of 165\ts pc (the cool plus warm
components) between 4 and 8\ts kpc, and it increases considerably
beyond the Solar circle (Henderson et al.\ 1982). Near the Sun the
$\HI$ scale height is about 200\ts pc, a factor of 3 smaller than the
scale height of the ionized gas. Assuming this ratio to be constant
for $r\ge r_{\sun}$, and taking the solar-neighborhood values at a
radius of 9\ts kpc in M51, we found the scale heights of the thermal
gas for the outer rings as given in Table~2. A sketch of M51
illustrating the spatial distribution of the various components is
shown in Fig.~2.

The filling factor of the free electrons in the thermal disk of
M51 was also adopted from the Solar neighborhood. It was
calculated by comparing the flux density of the thermal radio
emission with that expected from the disk of thermal electrons
(Berkhuijsen, in preparation).The thermal flux density at 2.8\ts cm
of the Solar neighborhood, scaled to the distance of M51 and seen
with a beam width of $75\arcsec$, is $S_{2.8} = 0.35\pm
0.10$~mJy/beam. Using Eq.~(1), with $C$ taken for the distance of M51,
$\langle\nelectron\rangle = 0.035\pm 0.005\cmcube$ and $\hth
=600$\ts pc (as applicable near the Sun) we obtained $f=0.075\pm
0.030$. This value is in good agreement with other estimates, being
halfway between the filling factor of the diffuse warm gas ($f>0.2$,
Reynolds 1991b) and that of giant $\HII$ regions ($f\simeq 0.01$,
G\"usten \& Mezger 1983).

As no information is available on how $f$ varies with $r$ in the
Milky Way, we adopted $f=0.075$ for all 4 rings in M51.

\titleb{ Depolarization }

In Sect.~3.3.1 and 3.3.2 of the electronic version we discuss
wavelength-independent depolarization and Faraday depolarization
outside the disk of M51. Based on that discussion, we assume that the
Faraday depolarization occurs entirely in the thermal disk of M51.
In order to take out the effect of the wavelength-independent
depolarization we shall use the relative depolarization between
$\lambda=20.5\cm$ and $\lambda=2.8\cm$, denoted as
$\DPT=P_{20.5}/P_{2.8}$. As Faraday effects at $\lambda2.8\cm$ are
negligible, this ratio is essentially a measure of the Faraday
depolarization at the longer wavelengths. The observed values are
given in Table~3.

\begtabfull
\tabcap{3}{
The observed relative depolarization between
$\lambda20.5\cm$ and $\lambda2.8\cm$, $\DPT$, the
vertical extent $\Delta z$ of the upper layer of  the thermal
disk visible in polarized emission
at $\lambda\lambda18.0/20.5\cm$, and the
corresponding values of $\xi\disk$ and $\xi\halo$ at $\lambda =
20.5\cm$ calculated from Eq.~(7).
}
\halign{
#\hfil&\qquad\hfil#\hfil
        &\qquad\hfil#\hfil&\qquad\hfil#\hfil&\qquad\hfil#\hfil \cr
\noalign{\hrule\medskip}
\hfil$r$ &$\DPT$
        &$\Delta z$ &$\xi\disk$ &$\xi\halo$    \cr
\hfil [kpc] & &[pc]  & & \cr
\noalign{\medskip\hrule\medskip}
\pheins 3--\pheins 6 &$0.38$        &\pheins200  &0.26  &0.96   \cr
\pheins 6--\pheins 9 &$0.37$        &\pheins300  &0.26  &0.92   \cr
\pheins 9--12        &$0.70$        &2000        &0.73  &0    \cr
12--15               &$0.94$        &4000        &1.00 &0    \cr
\noalign{\medskip\hrule}
}
\endtab

The Faraday depolarization in the disk is caused by both
differential Faraday rotation and internal Faraday dispersion.
Using results of Burn (1966) with the values in Tables 1, 2 and 3 we
find that in the disk of M51 each of these effects is strong
enough to significantly depolarize the emission at $\lambda20/18\cm$.
Due to internal Faraday dispersion, only an upper layer of the disk
is visible. As we estimate below, this layer is only about
200--300\ts pc deep at $r=3$--9\ts kpc. It can be easily seen that
depolarization due to differential Faraday rotation across this depth
is relatively weak.

The fact that only polarized emission from an upper layer is observed
is evident from the much smaller rotation measures observed between
$\lambda\lambda$20.5 and 18.0\ts cm (Horellou 1990) than between
$\lambda\lambda$6.3 and 2.8\ts cm (Neininger 1992b), at which
wavelenghts Faraday rotation is negligible. For the two inner rings
$\RM (20.5/18.0) \simeq 0.25\times \RM (6.3/2.8)$. As the disk is
transparent at short wavelengths, this indicates directly that only a
part of it is seen in polarized emission at $\lambda\approx20\cm$.
The visible depth in the disk is then estimated as $\Delta z\simeq
0.25\times2\hth =200$--300\ts pc for $r=3$--9\ts kpc. In the case
of field reversals in the part of the disk invisible at
$\lambda\lambda$20.5, 18.5~cm this value is an upper limit to $\dz$.
Horellou et al. (1992), using different arguments, also concluded
that at $\lambda\lambda 20.5,18.0$ only the upper part of the
polarized disk is observed, and Beck (1991) found the same for
NGC~6946.

As the data at $\lambda\lambda 6.3,2.8\cm$ are not complete for the
two outer rings, we cannot make the above comparison for $r=9$--15~kpc.
Instead we propose the following estimate of the {\it minimum visible
depth.\/} Let us define $\dz$ as the depth in the {\it thermal disk\/}
from which polarized emission is observed (see Fig.~2).
Then the layer in the synchrotron disk, which produces the observed
polarized emission at $\lambda$20.5\ts cm, has the thickness
$(h_{20.5}-\hth)+\dz$. If no Faraday depolarization occurred in the
visible layer, then the fraction $\DPT$ of the polarized emission
at $\lambda20.0\cm$ would come from a layer $\DPT\times2h_{20.5}$
deep. Since this depth must be equal to the former value, this yields
$$
\dz = h_{20.5}(2\,\DPT-1)+\hth\;.
\eqno(4)
$$
As some depolarization actually occurs within $\dz$, the true
visible depth must be larger than this. The values of $\dz$ thus
calculated are given in Table~3. These values are remarkably close to
the upper limits derived above from the $\RM$s observed in the two
wavelength ranges. We note that in the radial range 12--15\ts kpc the
thermal disk is completely transparent to polarized emission at
$\lambda\lambda18.0/20.5\cm$.

\titlec{Qualitative analysis of Faraday rotation in a double-layer system}
The polarization angle of the polarized emission is given by
$$
\psi = \RMfg\, \lambda ^2 + \RM\, \lambda ^2 + \psi _0\;,
\eqno(5)
$$
where $\RMfg$ is the foreground Faraday rotation measure produced
mainly within the Milky Way. \RM\ is the intrinsic Faraday rotation
measure produced by the magnetic field within the galaxy considered;
$\lambda$ is the wavelength and $\psi _0$ is the intrinsic
polarization angle.

In order to distinguish between the contributions of the disk and the
halo (see Sect.~4.1), we write
$$
\RM=\xi\disk\,\RM\disk+\xi\halo\,\RM\halo\;,
\eqno(6)
$$
where $\RM\disk$ and $\RM\halo$ are the Faraday rotation measures
produced across the disk and the halo, respectively, if both are
fully transparent to polarized emission. $\RM\disk=0.81
\langle\nelectron\rangle\disk B_\parallel\disk \hth$ is defined here
in terms of the {\it disk scale height\/}. (The full thickness of
the disk is $2\hth$.) However, it is more convenient to define
$\RM\halo = 0.81\langle\nelectron\rangle\halo
B_\parallel\halo Z$ in terms of the vertical extent of the halo, $Z$,
i.e., the distance along $z$ between $z=\hth$ and the upper boundary of
the halo (see Fig.~2). $B_\parallel$ is the line-of-sight component of
the regular magnetic field. Since not the whole disk (or even halo)
may be visible in polarized emission at a given wavelength, we
introduce factors $\xi\disk$ and $\xi\halo$. As follows from above,
$\xi\disk$ and $\xi\halo$ depend on the wavelength. We assume
that both $\xi\disk$ and $\xi\halo$ are the same at $\lambda2.8$ and
6.2\ts cm, and also at $\lambda18.0$ and 20.5\ts cm.

We can see from Tables 1 and 2 that the synchrotron disk at
$\lambda\lambda2.8$ and 6.2\ts cm is about as thick as or thinner
than the thermal one. Thus there is only little synchrotron emission
originating in the halo and the halo magnetic field can be detected
mainly via Faraday rotation in the near half. At $\lambda\lambda18.0$
and 20.5\ts cm, where $\hsyn>\hth$, the disk is not transparent to
polarized emission at $r<10\kpc$ where the halo is present.  As a
result, it is impossible to determine the structure of the magnetic
field in the part of the halo lying beyond the thermal disk from
observations of the intrinsic polarized emission.

Now we express $\xi\disk$ and $\xi\halo$ in terms of the scale
heights of the thermal and synchrotron disk, $\hth$ and $\hsyn$,
and $\dz$ in a given wavelength range. One should take into account
that, if synchrotron emission and Faraday rotation occur in the same
region, the observed Faraday rotation measure of a transparent layer
is equal to $\half\int B_\parallel\nelectron\,dL$, whereas that
produced in a foreground Faraday screen (i.e., a magneto-ionic layer
devoid of relativistic electrons) is $\int
B_\parallel\nelectron\,dL$. Assume that $\hsyn\geq\hth$, which
inequality is true in the case of M51 for $\lambda\approx20\cm$. The
Faraday rotation measure observed from the disk is given by
$\half\RM\disk\dz/\hth$. The contribution of the halo to the observed
Faraday rotation measure is  $\half\RM\halo (\hsyn-\hth)/Z +
\RM\halo[Z-(\hsyn-\hth)]/Z$, where the first term is due to the
synchrotron-emitting region, and the second one is the contribution
of the rest of the halo which acts as a foreground Faraday screen.
Thus, we have
$$
\xi\disk=\half{{\dz}\over\hth}\;,  \quad
\xi\halo=1-\half{{\hsyn-\hth}\over Z}\quad\hbox{\rm for}\ Z>0\;;
\eqno(7)
$$
$\xi\halo$ is undefined for $Z=0$ whereas $\RM\halo=0$ in
this case. Expressions for $\dz$ applicable in other cases can be
found in Sect.~3.3.3 of the electronic version of the paper.

At $\lambda = 2.8/6.2 \cm$ the galaxy is transparent, so that $\dz =
2\hth$ and we obtain $\xi\disk=1$; furthermore, $\xi\halo\approx1$
because at these wavelengths $\hsyn$ differs insignificantly
from $\hth$ for $r=3$--9\ts kpc and $\hsyn<\hth$
for $r=9$--12\ts kpc. However, $\xi\disk$ strongly differs from unity
at $\lambda = 18.0/20.5 \cm$. The dependence of $\xi\disk$
and $\xi\halo$ on $\lambda$ is due to the $\lambda$-dependence of
$\dz$ and $\hsyn$. The values of $\dz$, $\xi\disk$ and  $\xi\halo$
given in Table~3 refer to $\lambda\lambda18.0/20.5\cm$.
The halo is transparent for polarized emission at all the wavelengths
considered, and $\xi\halo$ differs from unity only because some
synchrotron emission originates within the halo (at $\hsyn\leq
z\leq\hth$), whereas the remaining part of the halo acts as a
foreground screen.

\titlea{Recognition of magnetic field patterns}
When fitting the observed distribution of the polarization angle
$\psi $ at a given wavelength $\lambda$ we adopt for each layer the
following truncated Fourier representation for the cylindrical
components of the regular magnetic field ${\vec B}_{\rm reg} =
(B_{r}, B_\theta , B_{z})$:
$$
\eqalign{
B_{r} &= B_0 \sin p_0 + B_1\sin p_1 \cos (\theta - \beta)\;, \cr
B_\theta &= B_0\cos p_0 + B_1 \cos p_1 \cos (\theta - \beta)\;, \cr
B_{z} &= B_{z0} + B_{z1}\cos (\theta - \beta_{z})\;, \cr
}
\eqno(8)
$$
where $B_0$ and $B_{z0}$ are the strengths of the horizontal
(parallel to the galactic plane) and vertical (perpendicular to the
plane) components of the $m=0$ mode, respectively, $B_1$ and $B_{z1}$
are those of the $m=1$ mode, $p_0$ and $p_1$ are the pitch
angles, and $\beta$ and $\beta_{z}$ are the azimuthal angles at which
the corresponding non-axisymmetric components are maximum.
In Eq.~(8) only the two lowest modes have been retained; this proves
to be sufficient to fit the available data. The magnetic pitch angle
is the small angle measured from the magnetic field vector to the
tangent of the local circumference. It is positive (negative) if the
magnetic field spiral opens counterclockwise (clockwise). We note
that the magnetic field direction can be either inwards or outwards
along the spiral. In the case of M51 a negative pitch angle
corresponds to a trailing spiral.

The intrinsic polarization angle $\psi _0$ in Eq.~(5) is
determined by the transverse component of the magnetic field. A
suitable expression relating $\psi _0$ to the magnetic field of
the form given in Eq.~(8) was given by Sokoloff et al.\ (1992); its
detailed derivation can be found in Appendix~A of the electronic
version -- see Eq.~(A3). As the synchrotron emissivity in the halo is
significantly weaker than in the disk, we assume that the intrinsic
polarization angle depends solely on the field in the disk.

We should emphasize that we analyze simultaneously and consistently
the longitudinal and transverse (with respect to the line of sight)
components of the magnetic field which manifest themselves through
\RM\ and $\psi _0$, respectively.  Previous work either considered
solely the Faraday rotation measures between pairs of wavelengths or
simplified the model by supposing that $\psi _0 = {\rm const}$ (see
Sokoloff et al.\ 1992).  In both cases only the line-of-sight
magnetic field could be recovered from observations.  Attempts to
extract additional information about the transverse component of the
magnetic field from an independent analysis of ``magnetic pitch
angles,'' or $\psi_0$, from total and polarized intensity, etc.\
often led to results that were inconsistent with those obtained from
the \RM\ analysis.  Here for the first time we propose a
{\it consistent\/} three-dimensional model of the regular magnetic
field observed in a galaxy.

The fitting procedure and the estimation of uncertainties of the
fitted parameters are discussed in detail in Sect. 4.2, 4.4, and
Appendix~B of the electronic version of the paper.

In order to obtain satisfactory fits to the data, we calculated the
residual $S$ defined in Eq.~(B1), that characterizes the deviation of
the fit from the measured points, found its minimum with respect to
the above fit parameters, and then employed the $\chi^2$ and Fisher
statistical tests briefly discussed in Appendix~B to assess the
reliability of the fit. The $\chi^2$ statistical test, Eq.~(B2),
ensures that the fit is close enough to the measured points with
allowance for their weights equal to $\sigma_{ni}^{-2}$ with
$\sigma_{ni}$ the standard deviation of $\psi_{ni}$ (see Sect.~2).
The Fisher test, Eq.~(B3), was then applied to verify that the
quality of the fit is the same at all individual wavelengths (see
Sokoloff et al.\ 1992 for details).

\titleb{Statistical evidence for a magneto-ionic halo in M51}
Fits for all wavelengths obtained with one transparent magneto-ionic
layer were generally inconsistent with the Fisher test at $r<9\kpc$.
Therefore, we tried fits to the polarization angle distributions for
$\lambda\lambda$2.8/6.2~cm and $\lambda\lambda$18.0/20.5~cm
separately. The magnetic patterns
revealed in the two wavelength ranges in the inner two rings turned
out to be very different from each other. For example, the values of
$B_0$ obtained at the shorter and longer wavelengths had
different signs. Apparently the only plausible explanation for
this is that different regions are sampled in the two wavelength
ranges and that the magnetic field has completely different
configurations in them. Moreover, $B_{\parallel}$ changes sign either
in the upper part of the disk $\dz$ or in a layer above $z=\hth$.
Since it is very unlikely that the regular magnetic
field may have such a complicated vertical structure within the
upper part of the disk, we concluded that the change of direction of
$B_{\parallel}$ suggests the existence of one more extended
component in the galaxy. Thus, at $r\leq9\kpc$, at least two {\it
extended\/} components of the magneto-ionic medium in M51 are
present, namely the disk and the halo. Then the observed Faraday
rotation at the short wavelengths is dominated by the magnetic field
in the disk whereas that at the long wavelengths is mainly determined
by the field in the halo.

It is clear that separate fits for different wavelength ranges cannot
be physically satisfactory. Therefore all the fits discussed below
were obtained from all four wavelengths {\it simultaneously\/} using
the values of $\dz$ in Table~3. For the inner rings, $r=3$--9\ts
kpc, we applied a double-layer model of the magneotionic medium using
the values of $\dz$ in Table~3, whereas for the radial range
9--15\ts kpc the fits were performed for a single-layer model without
distinguishing the disk and halo contributions to $\RM$ (that is, we put
$B_i\halo=0$ there).

Evidence of a halo in M51 from X-ray observations is discussed in
Sect.~4.3 of the electronic version of the paper. We adopted an
elliptical shape for the halo with a height of 6\ts kpc above the
midplane near the center (see Fig.~2) and assumed an electron density
of $3\times10^{-3}\cmcube$ as in the Milky Way.

\titleb{Results of the fitting}
The results of the fitting are presented in detail in Sect.~4.4 of the
electronic version of the paper. Here we combine parts of Tables~4 and
5 and we only mention that the regular magnetic field
in the disk is shown to be a superposition of axisymmetric and
bisymmetric modes with a slight predominance of the latter. The field in
the halo appears to be axisymmetric. An example of the fits is shown in
Fig.~3 for $r=3$--6\ts kpc. Similar figures for the other rings can be
found in the electronic version.

\begtabfullwid
\tabcap{$\!$s 4 and 5 (combined)}{A fitted model for the regular
magnetic field in M51$^{\,\rm (a)}$}
\halign{
#\hfill $\;$  &#\hfill\qquad
        &# \hfill\quad  &# \hfill\qquad
        &# \hfill\quad  &# \hfill\qquad
        &# \hfill\qquad  &# \hfill
                                           \cr
\noalign{\hrule\smallskip}
$r$  &[kpc]
            &\multispan2\hfill  3--6             \qquad\quad\hfill
            &\multispan2 \hfill 6--9             \qquad\quad\hfill
            &  9--12                       
            &  12--15                
                                  \cr
     &
                & \multispan2 \hskip-0.5em\hrulefill \qquad\quad
                & \multispan2 \hskip-0.5em\hrulefill \qquad\quad
                & \multispan1 \hskip-0.5em\hrulefill \qquad\quad
                & \multispan1 \hskip-0.5em\hrulefill         \cr
\noalign{\smallskip}
         &
                &Disk           &Halo
                &Disk           &Halo
                &Disk
                &Disk
                                    \cr
\noalign{\medskip\hrule\medskip}
$\RMfg$ &[$\!\radm $]
                &\multispan2 \hfill $+3\ers{2}$ \qquad\quad \hfill
                &\multispan2 \hfill $+8\ers{5}$ \qquad\quad \hfill
                &$+9\ers{4}$ 
                &$+5^{\,\rm (b)}$  
                                        \cr
\noalign{\medskip}
$B_0$  &[$\!\mkG$]
                &$-4.8\err{+1.3}{-2.6}$   &$+3.4\ers{0.8}$
                &$-5.2\err{+3.0}{-5.8}$   &$+1.0\ers{2.6}$
                &$-1.6\err{+1.1}{-7.6}$
                &$-3.2\err{+2.4}{-31.3}$           \cr
\noalign{\medskip}
$p_0$   &[deg]
                &$-12\ers{1}$      &$-23\ers{13}$
                &$-12\ers{2}$      &$-23\ers{180}$
                &$-6\ers{3}$
                &$+18\err{+4}{-7}$         \cr
\noalign{\medskip}
$B_1$  &[$\!\mkG$]
                &$-6.9\err{+2.1}{-3.2}$   &$\cdots$
                &$-4.9\err{+4.8}{-2.2}$  &$\cdots$
                &$-1.4\err{+1.0}{-6.8}$
                &$-3.2\err{+2.4}{-28.9}$               \cr
\noalign{\medskip}
$p_1$   &[deg]
                &$-11\ers{2}$           &$\cdots$
                &$-10\err{+10}{-3}$     &$\cdots$
                &$-3\ers{3}$
                &$+32\err{+5}{-13}$         \cr
\noalign{\medskip}
$\beta$ &[deg]
                &$+176\ers{6}$         &$\cdots$
                &$+68\err{+34}{-68}$   &$\cdots$
                &$+58\ers{6}$
                &$-63\err{+7}{-10}$        \cr
\noalign{\medskip}
$B_{z0}$  &[$\!\mkG$]
                &$\cdots$               &$\cdots$
                &$\cdots$               &$\cdots$
                &$\cdots$
                &$+1.9\err{+14.2}{-1.4}$               \cr
\noalign{\medskip}
$B_{z1}$  &[$\!\mkG$]
                &$\cdots$               &$\cdots$
                &$\cdots$               &$\cdots$
                &$\cdots$
                &$+0.5\err{+5.7}{-0.5}$               \cr
\noalign{\medskip}
$\beta_{z}$  &[deg]
                &$\cdots$               &$\cdots$
                &$\cdots$               &$\cdots$
                &$\cdots$
                &$+186\ers{8}$                \cr
\noalign{\medskip}
$\overline{B}_{\rm reg}$  &[$\!\mkG$]
                &$6.9\err{+3.4}{-2.0}$        &$3.4\ers{0.8}$
                &$6.3\err{+5.8}{-4.0}$        &$1.0\err{+2.6}{-1.0}$
                &$1.9\err{+9.0}{-1.3}$
                &$4.4\err{+40.2}{-3.2}$                \cr
\noalign{\medskip\hrule\medskip}
\noalign {
\parshape=1 3pt 459.7pt
\noindent
{\petit {\bf Notes:}
\newline ${}^{\rm (a)}\,$
Dots mean that the corresponding parameter is insignificant.
Errors are 2$\sigma$ Gaussian deviations.
\newline ${}^{\rm (b)}\,$
$\RMfg$ for 12--15\ts kpc was fixed at the given value (see Sect.\
4.4.4 of the electronic version).
}}\noalign{\medskip\hrule}}
\endtab

\begfigwid 9.9cm    
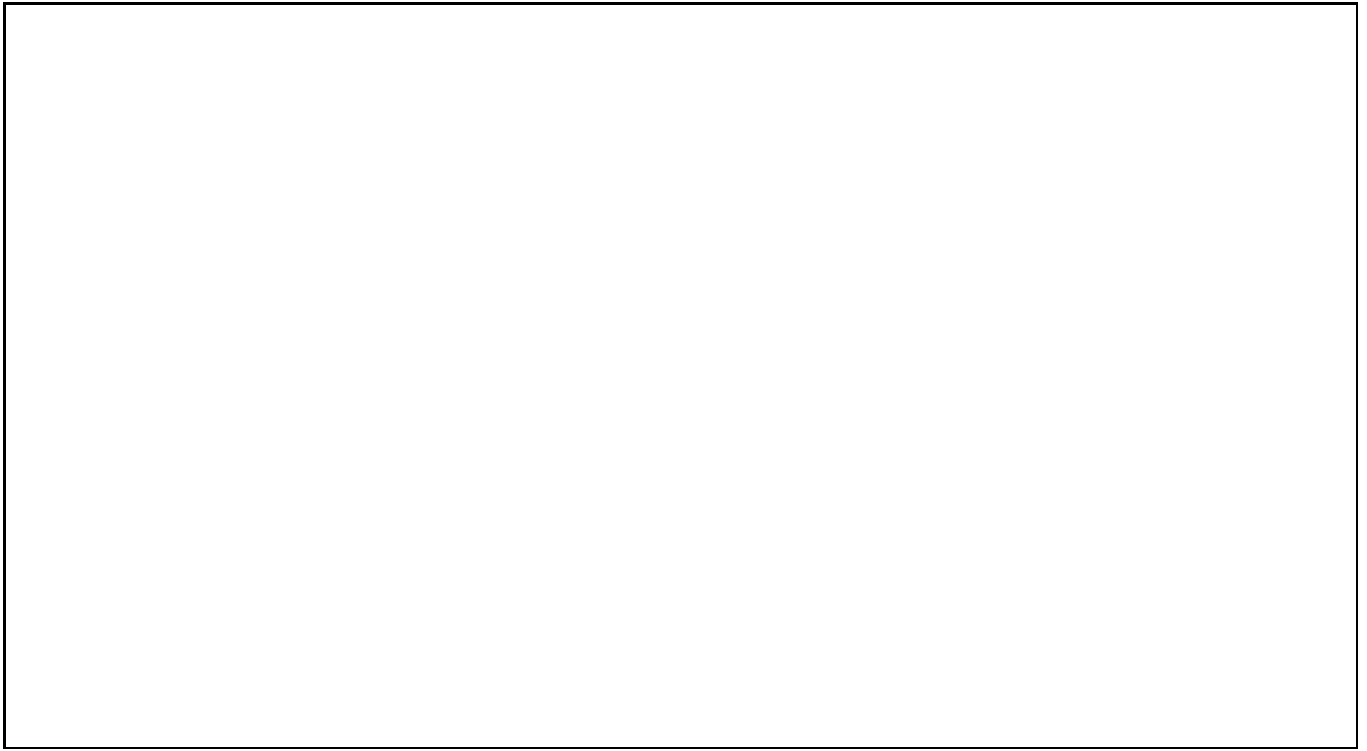
\figure{3}{The polarization angles (dots with error bars, measured
from the local radial direction in the plane of M51) as a function
of azimuthal angle in the galactic plane and fits (solid lines) for
the ring $3\leq r \leq 6\kpc$ at (a) $\lambda2.8$, (b) $\lambda6.2$,
(c) $\lambda18.0$ and (d) $\lambda20.5$\ts cm. Error bars show the
1$\sigma$ errors of the measurements. Dashed lines indicate the
foreground levels, $\RMfg\lambda^2$ }
\endfig

\titlea{Global structure of the regular magnetic field}
In this section we derive the strength and the direction of the regular
magnetic field, and we discuss the global properties of these parameters.

The results are compiled in Tables~4 and 5.
A note of caution is appropriate here: the resulting amplitudes
$B_i$ were obtained assuming that $\langle\nelectron\rangle$ is
independent of azimuthal angle. For each ring we also give the
strength of the regular magnetic field averaged over the azimuth,
$\overline{B}_{\rm reg}$. Uncertainties were calculated with
allowance for errors in $\langle\nelectron\rangle$ too. The large
errors, especially at 12--15\ts kpc, are mainly due to the errors in
$\langle\nelectron\rangle$.

\putattop
\begfig 5.5 cm  
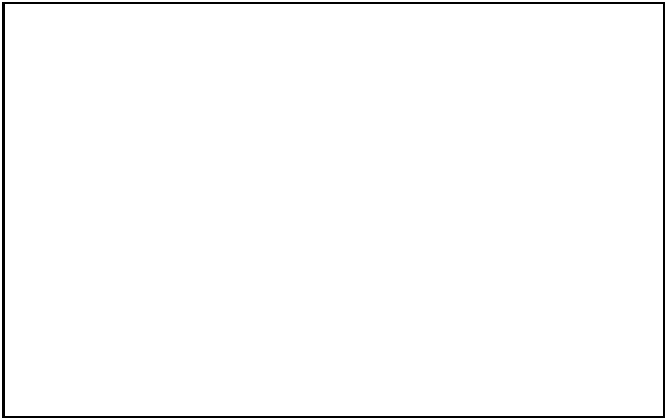
\figure{7}{The radial variation of the strength of the regular
magnetic field obtained from our fits and averaged in the rings
(circles with error bars), the total magnetic field obtained from
the total intensity of the nonthermal emission assuming energy
equipartition between cosmic-ray particles and magnetic field
(dashed) and its regular component obtained using the
observed degree of polarization (solid).}
\endfig

In Fig.~7 we show the radial variation of $\overline{B}_{\rm reg}$
and that of the total and regular magnetic field strengths, $B$ and
$\B$, obtained from the total nonthermal emission and the observed
degree of polarization as described in Sect.~3.1 and given in Table~1.
We note that for each ring there is a close agreement between the
two values of the regular magnetic field which were obtained from
completely {\it independent\/} physical parameters and methods.
Between $r=3$ and 15~kpc the exponential radial scale length is
$14.3\pm 6.0\kpc$.

As can be seen from Tables~4 and 5 the magnetic fields in the halo
and in the disk inside $r=12$\ts kpc are horizontal. In the
halo we have $B_0\halo > 0$, $p_0\halo<0$ and $B_1\halo \simeq 0$,
which means that the radial component of the regular magnetic field
is directed {\it inwards,\/} with the azimuthal component directed
counterclockwise; that is, $B_r\halo<0$ and $B_\theta\halo>0$ --
see Eq.~(8). Meanwhile, in the
disk we have $B_0\disk\approx B_1\disk$ and both are negative
together with the pitch angles. Therefore, for $r=3$--9\ts kpc the
radial field in the disk is directed {\it outwards\/}, with
$B_\theta$ directed clockwise at almost all $\theta$. We conclude
that {\it the regular magnetic fields in the disk and the halo
have almost opposite directions\/} everywhere within $r=9$\ts kpc
except in the northwestern part of the ring at $r=3$--6\ts kpc.

\begfigwid 9.2 cm
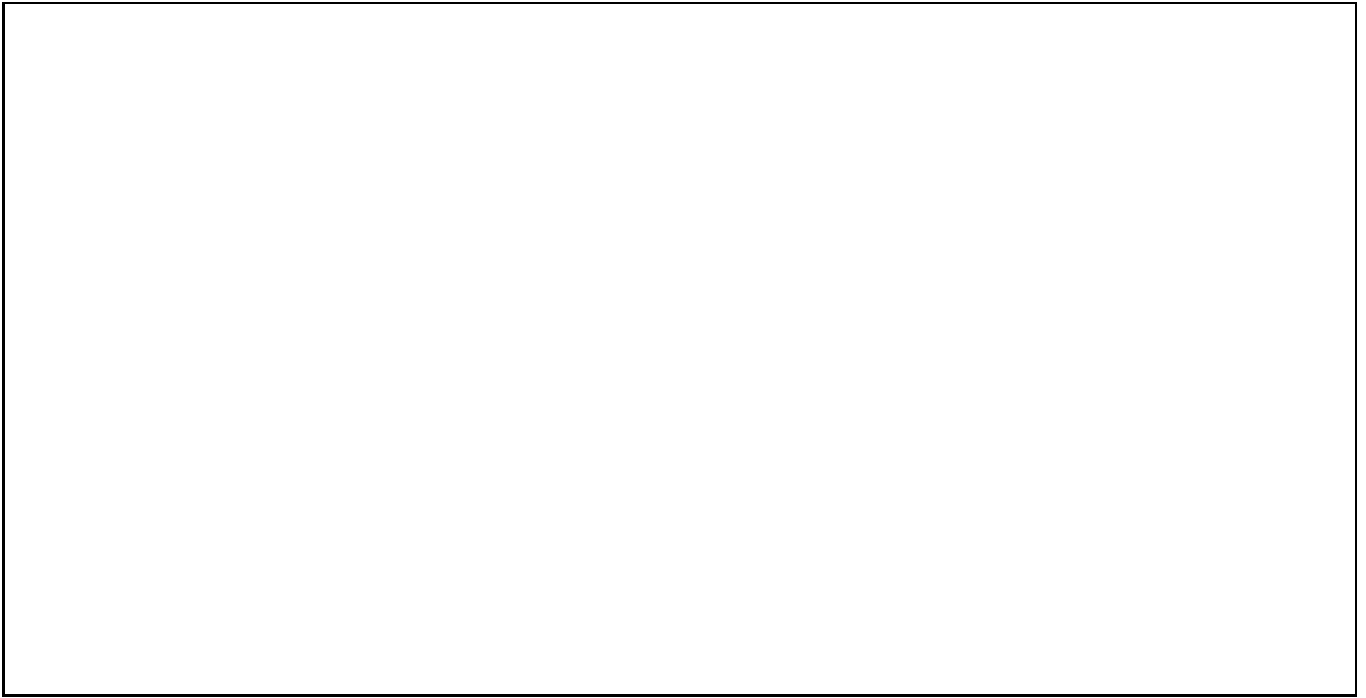
\figure{8}{Directions of the horizontal regular magnetic field
in the disk (a) and halo (b) of M51 according to the fits presented
in Tables~4 and 5. For clarity we scaled the vectors as follows: they
are proportional to $\R$ ($= \B \langle\nelectron\rangle \hth$) in the
inner two rings in the disk, to $3\R$ in the two outer rings in the
disk, to $2.5\R\halo$ $(= \B \langle\nelectron\rangle Z)$ for 3--6\ts
kpc in the halo, and to $5\R\halo$ for 6--9\ts kpc
in the halo. The vertical component at $r=12$--15\ts kpc was not
included. The vectors are shown superimposed onto an optical picture
of M51. The sectors and rings used are indicated.
}
\endfig

In Fig.~8 the direction of $\vec{B}_{\rm reg}$ in each sector is
shown for the disk and the halo separately. The length of the vectors
is proportional to $\B \langle\nelectron\rangle \hth$ for the disk and
$\B \langle\nelectron\rangle Z$ for the halo with scaling factors
specified in the caption.

In the outer ring, 12--15\ts kpc, the magnetic field structure is
distorted. The values of $p_0$ and $p_1$ differ considerably from
those for $r\leq9\kpc$ and are even positive. Inspection of the
polarization map in Fig.~1a confirms that in the northern part
the magnetic pattern at these radii is plagued by strong distortions
still having a rather large spatial scale. Only in the ring 12--15~kpc
the magnetic field has a weak vertical component.

\titlea{Discussion}
\titleb{The magnetic field in the halo}
The available polarization measurements performed at the two pairs of
widely separated wavelengths allowed us to determine the magnetic field
structure in two regions along the line of sight. In the text above
we called these regions the disk and the halo. This usage was justified
in Sect.~4.1.

Our results represent the first indication of a magneto-ionic halo in a
galaxy seen nearly {\it face-on\/}. The detection of a radio halo in an
edge-on galaxy is a difficult observational problem, even more so the
determination of the magnetic field structure. In the galaxies seen
nearly face-on some of the difficulties are alleviated. First, the
halo and its magnetic field are illuminated by a strong background
source of polarized emission, the disk. Second, the polarization
measurements over the entire disk can be used to reveal the global
azimuthal structure of the field in the halo as it was done in the
present paper. It is important to note that the magnetic field in M51
has different structures in the disk and the halo. If the field
structures were similar to each other, the detection of the
magneto-ionic halo might be difficult.

We showed that the field in the halo of M51 is predominantly
horizontal as in the halos of NGC~891 and NGC~253 (Hummel et al.\
1991b; Sukumar \& Allen 1991; Beck et al.\ 1994).

According to our fits we estimate the halo radius to be about
10\ts kpc.  This estimate agrees with the data on X-ray emission from
M51 which also indicate a halo radius of about 10\ts kpc (Ehle et
al.\ 1995).

With the values of $\langle\nelectron\rangle$ and $Z$ from Table~2, the
estimated strength of the {\it regular\/} magnetic field in the halo
decreases from about 3$\mkG$ at the radial distance of 3--6$\kpc$ to about
1$\mkG$ at $r=6$--9\ts kpc. The field is basically axisymmetric. The upper
limits on the $m=1$ mode in the halo are estimated from our fits as
$|B_1\halo|\la1\mkG$ and $\la 3\mkG$ for $r=3$--6 and 6--9\ts kpc,
respectively.

It is interesting to compare the values of the regular magnetic
field in the halo with the upper limit on the total magnetic field
strength estimated from the equilibrium between thermal and
magnetic energy densities in the X-ray emitting gas
(Ehle et al.\ 1995). With $\langle\nelectron\rangle=0.003\cmcube$ and
a volume filling factor of 0.8, their results yield $\H<7\mkG$. Our
results are consistent with this limit. If the true total field
strength is close to the above upper limit, the turbulent field
in the halo has a strength of about $6\mkG$ exceeding that of
the regular magnetic field.

The global field directions are in general {\it opposite in the disk
and the halo\/} of M51. This implies that the regular
magnetic field in the halo cannot be simply advected from the disk.
Such reversals appear in the dynamo theory for galactic halos
(Ruzmaikin et al.\ 1988, Sect.\ VIII.1; Sokoloff \& Shukurov 1990;
Brandenburg et al.\ 1992) and could be due to the topological pumping
of magnetic field by a galactic fountain flow (Brandenburg et al.\
1995). Moreover, the dominance of the axisymmetric field in the halo
is also consistent with the mean-field dynamo theory which predicts
that non-axisymmetric magnetic modes can be maintained only in a thin
galactic disk and most likely decay in a quasi-spherical halo (see
Ruzmaikin et al. 1988). We cannot say anything about the parity of
the halo field with respect to the midplane because the galaxy is not
transparent at $\lambda \approx 20 \cm$ in the rings where the halo
is present and, in addition, the synchrotron emission from the halo
is negligible.

\titleb{The azimuthal structure of the field}
The azimuthal distributions of polarization angle in M51 seen
over the radial range $r=3$--15\ts kpc are successfully represented
by a superposition of only {\it two\/} azimuthal modes
of the large-scale magnetic field, $m=0$ and $m=1$ in the disk.
Even though we restrain ourselves from identifying these magnetic
harmonics with dynamo-generated axisymmetric and bisymmetric modes
before a more careful theoretical analysis has been made, we mention
that dynamo theory also predicts that the two leading azimuthal modes
$m=0$ and $m=1$ typically dominate in spiral galaxies. Furthermore, it
follows from the dynamo theory that non-axisymmetric magnetic
structures should be more often a superposition of the two azimuthal
modes than a purely bisymmetric mode (Ruzmaikin et al.\ 1988, p.~231;
Beck et al.\ 1996).  A similar superposition of modes, but with a
dominance of the bisymmetric mode, was found earlier in M81 by
Sokoloff et al.\ (1992) (see also Krause et al. 1989b). In M31 the
axisymmetric magnetic mode is dominant (Ruzmaikin et al.\ 1990), and
for NGC~6946 a superposition of $m=0$ and $m=2$ magnetic modes was
suggested by Beck \& Hoernes (1996).

Of course our results do not imply that higher azimuthal magnetic
modes are not present in M51, but only that the accuracy of
the available observations is insufficient to reveal them. One can
expect that the amplitudes of the harmonics with $m\geq 2$ are
considerably smaller than those of the modes $m=0$ and 1.

Since the theory of the galactic mean field dynamo predicts an efficient
generation of the bisymmetric mode in M51 with the maximum of the
$m=1$ eigenmode at $r\simeq 2\kpc$ (Baryshnikova et al.\ 1987;
Krasheninnikova et al.\ 1989), we are tempted to identify the $m=1$
mode revealed for $3\leq r \leq 9\kpc$ with a bisymmetric field
generated by the dynamo.  This suggestion is confirmed by the
closeness of the pitch angles $p_0$ and $p_1$ of the $m=0$ and 1
modes to each other in the two innermost rings:  this is typical of
the dynamo-generated fields in a thin disk (Ruzmaikin et al.\ 1988).
This conclusion is also plausible for $r=9$--12\ts kpc.
We also note that a nonlinear dynamo model of Bykov et al.\ (1996)
predicts a mixture of magnetic modes, which is roughly similar to that
detected here, to be found in some vicinity of the corotation radius,
i.e.\ just near 6\ts kpc in M51.

The azimuthal modes inferred for the outermost ring can be  hardly
identified directly with the dynamo modes because the pitch angles of
individual modes are positive. These modes may be due to distortions
imposed by non-axisymmetric density and velocity distributions
possibly caused by the encounter with the companion galaxy NGC~5195
(Howard \& Byrd 1990). Concerning the total horizontal regular
magnetic field, its pitch angle is negative for
$70\degr\leq\theta\leq170\degr$ and positive in the rest of the
sectors. Inspection of polarization maps in Fig.~1 confirms that the
pattern of polarization angles at these radii in the northern part is
strongly distorted on a rather large scale.

We note that the regular magnetic field in the disk of M51 is directed
outwards, whereas those in IC~342 (Krause et al.\ 1989a), M31 and
NGC~6946 (Beck et al.\ 1996) are directed inwards.

\titleb{ Inner and outer spiral structure }
Inspection of Tables~4 and 5 shows that for the disk most of the
fitted parameters of the inner rings ($r<9$\ts kpc) differ
systematically from those of the outer rings. The phase angle $\beta$
varies by about $110\degr$ between the rings at 3--6\ts kpc and
6--9\ts kpc, and also between 9--12\ts kpc and 12--15\ts kpc, but not
between 6--9 and 9--12\ts kpc. The inner pattern is more coherent and
has stronger magnetic field than the outer pattern. Thus it seems
that the magnetic field structure in the outer rings is not a smooth
continuation of the structure in the inner rings, but that rather two
distinctly different magnetic field structures are present in M51.

This result is very interesting as Elmegreen et al.\ (1989) showed,
using optical plates, that M51 contains an inner and an outer spiral
structure which are overlapping between $r=6$\ts kpc and $r=8$\ts
kpc (see Sect.~6.3 in the electronic version for a more detailed
discussion). The outer spiral arms are thought to be material arms
driven by the companion, whereas the inner spiral arms are due to
density waves caused by the outer arms.

A discontinuity in the magnetic field pattern indicates that different
physical effects contribute to the field structure at
$r\la9\kpc$ and $r\ga9\kpc$. The discontinuity occurs at the radius
where the inner spiral structure ends and the outer spiral structure
becomes dominant. Therefore, the two magnetic field structures can be
physically connected with the inner and outer spiral patterns proposed
by Elmegreen et al.\ (1989).

The relatively strong magnetic field and its regular pattern in the inner
region are compatible with the idea of a dynamo acting under more or less
steady conditions. In the outer regions, where the spiral arms are produced
by a recent encounter with NGC~5195 about $10^8$~years ago (Howard \& Byrd
1990), the magnetic field may be a remnant of an older one disrupted by
the velocity perturbation.  Therefore it is understandable that the pitch
angles are irregular and the magnetic field is weak.

We conclude that the magnetic field pattern in the disk of M51
appears to be not one global structure, but consists of an inner
pattern associated with the inner spiral structure of density wave
arms and an outer pattern related to the outer spiral structure
of material arms. The interaction between magnetic fields and the
spiral patterns is not yet understood.

\titleb{Pitch angles of the magnetic field and of the spiral arms}

We compared the pitch angles of the magnetic field
in the disk derived from Eq.~(10) (see Sect.~6.4 in the electronic
version) with the pitch angles of the dust
lanes running along the inner edge of the optical spiral arms as
tabulated by Howard \& Byrd (1990). In each ring the optical pitch angles
were averaged in the same sectors as were used for the model fits (see
Fig.~8). The comparison was possible only for the inner two rings, as at
larger radii the measured optical pitch angles and the magnetic model
pitch angles have too few sectors in common.

Comparing the corresponding sectors we found general agreement
between optical and magnetic model pitch angles. For the ring 3--6\ts
kpc the mean of the optical pitch angles is $-15\degr\pm8\degr$ and
that of the magnetic pitch angles is $-11\degr\pm3\degr$, whereas for
the ring 6--9\ts kpc these values are $-13\degr\pm12\degr$ and
$-10\degr\pm8\degr$, respectively. The errors are one standard
deviation from the mean value and are due to intrinsic variation in
pitch angle in each ring. Although the agreement is quite good, we
note that the optical pitch angles show larger variations than those
of the fitted magnetic field.

Altogether, we conclude that on average the magnetic field inferred
from our fits is well aligned with the spiral arms, although local
misalignments may be considerable (see Fig.~8).

\bigskip
Section 6.5 {\it The origin of the vertical field\/} and also Appendices
A.~{\it The intrinsic polarization angle,\/} B.~{\it Statistical tests
and errors\/} and C~{\it Basic notation\/} can be found in the electronic
version of the paper.

\titlea{Conclusions}
1. The global magnetic pattern in M51 at $3\leq r\leq 15$\ts kpc can be
represented as a superposition of the two lowest azimuthal Fourier modes.

2. Our analysis indicates the existence of a magneto-ionic halo and shows
that the magnetic fields in the disk and the halo have different
configurations.

3. The radial extent of the halo is about 10\ts kpc, in agreement with
X-ray data. The halo field is horizontal and axisymmetric. The regular
magnetic fields in the halo and in the disk are spirals. {\it The field
directions along the spirals are generally opposite running inwards and
outwards in the halo and the disk, respectively.\/}

4. In the disk a superposition of axisymmetric ($m=0$) and bisymmetric
($m=1$) magnetic modes provides a satisfactory fit to the
observations. The $m=1$ mode slightly dominates at $3<r<6$\ts kpc and
the two modes have about equal amplitudes at $6<r<15$\ts kpc. The
field is predominantly horizontal between 3 and 12\ts kpc and has a
weak vertical component at $12<r<15$\ts kpc.  In the rings between 3
and 9\ts kpc the field structure in the disk is strongly
non-axisymmetric with a field maximum in the eastern part of
the galaxy and a weak field in the western part. Details are given in
Tables~4 and 5 and shown in Fig.~8.

5. The magnetic field pattern in the disk of M51 shows a discontinuity at
$r\simeq9$\ts kpc, the position at which the inner and the outer spiral
structure join (Elmegreen et al. 1989). The relatively strong, coherent
magnetic field in the inner rings occurs in the system of spiral arms
excited by density waves, whereas the weaker and partly distorted field in
the outer rings exists in the area of the material spiral arms produced by
the encounter with the companion.

6. The azimuthally averaged strength of the regular magnetic field obtained
for the disk decreases from about $7\mkG$ at radius 3--$6\kpc$ to about
$4\mkG$ at 12--15\ts kpc (see Fig.~7).  The strength of the regular field
in the halo decreases from $3\mkG$ at 3--6\ts kpc to zero beyond 9\ts kpc.
Details are given in Tables~4 and 5.

7. The azimuthal averages of the regular magnetic field strength in the
disk obtained from our fits are in good agreement with independent
estimates from the total synchrotron emission and the degree of
polarization. The radial scale length of the regular magnetic field is
$14\pm 6$~kpc.

8. We compared the pitch angles of the regular magnetic field obtained from
the fits with those of the dust lanes delineating spiral arms for the
sectors they have in common. Their mean values agree to within the errors.

\acknow{We are grateful to R.~Beck for numerous helpful discussions and
critical reading of the manuscript. A generous assistance of G.~Breuer is
gratefully acknowledged.  We thank S.~Sukumar for useful suggestions which
improved the presentation of results.  A.D.P., A.S.\ and D.D.S.\
acknowledge financial support of the Russian Foundation for Basic Research
under grants No.\ 93--02--3638, 95--02--03724 and 96--02--16252 and the
International Science Foundation under grant MNP000/300. A.D.P., A.S.\ and
D.D.S.\ are grateful to R.~Wielebinski for hospitality at the
Max-Planck-Institut f\"ur Radioastronomie (Bonn).  A.D.P., A.S., D.D.S.,
M.K., and N.N.\ thank the Deutsche Forschungsgemeinschaft and the Russian
Academy of Sciences for financial support during mutual visits.  }

\begref{References}
\ref Baryshnikova Y., Ruzmaikin A., Sokoloff D., Shukurov A., 1987, \AA\
     177, 27
\ref Beck R., 1991, \AA\ 251, 15
\ref Beck R., 1993. In: \potsdam , p. 283
\ref Beck R., Hoernes P., 1996, Nat 379, 47
\ref Beck R., Carilli C.L., Holdaway M.A., Klein U., 1994, \AA\ 292,
     409
\ref Beck R., Brandenburg A., Moss D., Shukurov A., Sokoloff D.,
        1996, ARAA 34, 153
\ref Berkhuijsen E.M., Golla G., Beck R., 1991. In: \leiden , p. 233
\ref Beuermann K., Kanbach G., Berkhuijsen E.M., 1985, \AA\ 153, 17
\ref Brandenburg A., Donner K.J., Moss D., Shukurov A., Sokoloff D.D.,
     Tuominen I., 1992, \AA\  259, 453
\ref Brandenburg A., Moss D., Shukurov A., 1995, MNRAS 276, 651
\ref Burn B.J., 1966, \MNRAS\ 133, 67
\ref Bykov A.A., Popov V.Yu., Shukurov A., Sokoloff D.D., 1996,
     MNRAS (submitted)
\ref Dickey J.M., Lockman F.J., 1990, \ARAA\ 28, 215
\ref Ehle M., Pietsch W., Beck R., 1995, \AA\ 295, 289
\ref Elmegreen B.G., Elmegreen D.M., Seiden P.E., 1989, ApJ 343, 602
\ref Ford H., Crane P., Jacoby G., Laurie D., 1985, ApJ 293, 132
\ref G\"usten R., Mezger, P.G., 1983, Vistas in Astronomy 26, 159
\ref Henderson A.P., Jackson P.D., Kerr F.J., 1982, \ApJ\ 263, 116
\ref Horellou C., 1990, Diploma Thesis, University of Bonn
\ref Horellou C., Beck R., Berkhuijsen E.M., Krause M., Klein U.,
     1992, \AA\ 265, 417
\ref Howard S., Byrd G.G., 1990, AJ 99, 1798
\ref Hummel E., Dahlem M., van der Hulst J.M., Sukumar S., 1991a, \AA\
     246, 10
\ref Hummel E., Beck R., Dahlem M., 1991b, \AA\ 248, 23
\ref Israel F.P., Habing H., de Jong T., 1973, \AA\ 27, 143
\ref Klein U., Wielebinski R., Beck R., 1984, \AA\ 135, 213
\ref Krasheninnikova Y., Ruzmaikin A., Sokoloff D., Shukurov A.,
     1989, \AA\ 213, 19
\ref Krause M., 1990. In: \heidel , p. 187
\ref Krause M., Beck R., Klein U., 1984, \AA\ 138, 385
\ref Krause M., Hummel E., Beck R., 1989a, \AA\ 217, 4
\ref Krause M., Beck R., Hummel E., 1989b, \AA\ 217, 17
\ref Mathewson D.S., van der Kruit P.C., Brouw W.N., 1972, \AA\ 17,
     408
\ref Mezger P.G., Henderson A.P., 1967, \ApJ\ 147, 471
\ref Neininger N., 1992a, \AA\ 263, 30
\ref Neininger N., 1992b, PhD Thesis, University of Bonn
\ref Neininger N., Horellou C., 1996. In: Roberge W.G., Whittet D.C.B.
     (eds.) Polarimetry of the Interstellar Medium. ASP Conf. Ser.
     Vol. 97, p.\ 592   
\ref Neininger N., Klein U., Beck R., Wielebinski R., 1991, Nat 352,
     781
\ref Neininger N., Horellou, C., Beck, R., Berkhuijsen, E., Krause, M.,
     Klein, U., 1993a. In: \potsdam , p. 313
\ref Neininger N., Beck R., Sukumar S., Allen R.J., 1993b, \AA\
     274, 687
\ref Reich P., Reich W., 1988, \AA\ 196, 211
\ref Reynolds R.J., 1991a. In: \leiden , p.~67
\ref Reynolds R.J., 1991b, \ApJ\ 372, L17
\ref Ruzmaikin A.A., Shukurov A.M., Sokoloff D.D., 1988, Magnetic Fields
     of Galaxies. Kluwer Acad.\ Publ., Dordrecht
\ref Ruzmaikin A., Sokoloff D., Shukurov A., Beck R., 1990, \AA\ 230, 284
\ref Sandage A., Tammann G.A., 1974, \ApJ\ 194, 559
\ref Segalovitz A., Shane W.W., de Bruyn A.G., 1976, \Nat\ 264, 222
\ref Sokoloff D., Shukurov A., 1990, Nat 347, 51
\ref Sokoloff D., Shukurov A., Krause M., 1992, \AA\ 264, 396
\ref Sukumar S, Allen R.J., 1991, \ApJ\ 382, 100
\ref Tosa M., Fujimoto M., 1978, \PASJ\ 30, 315
\ref Tully R.B., 1974, ApJS 27, 415
\endref

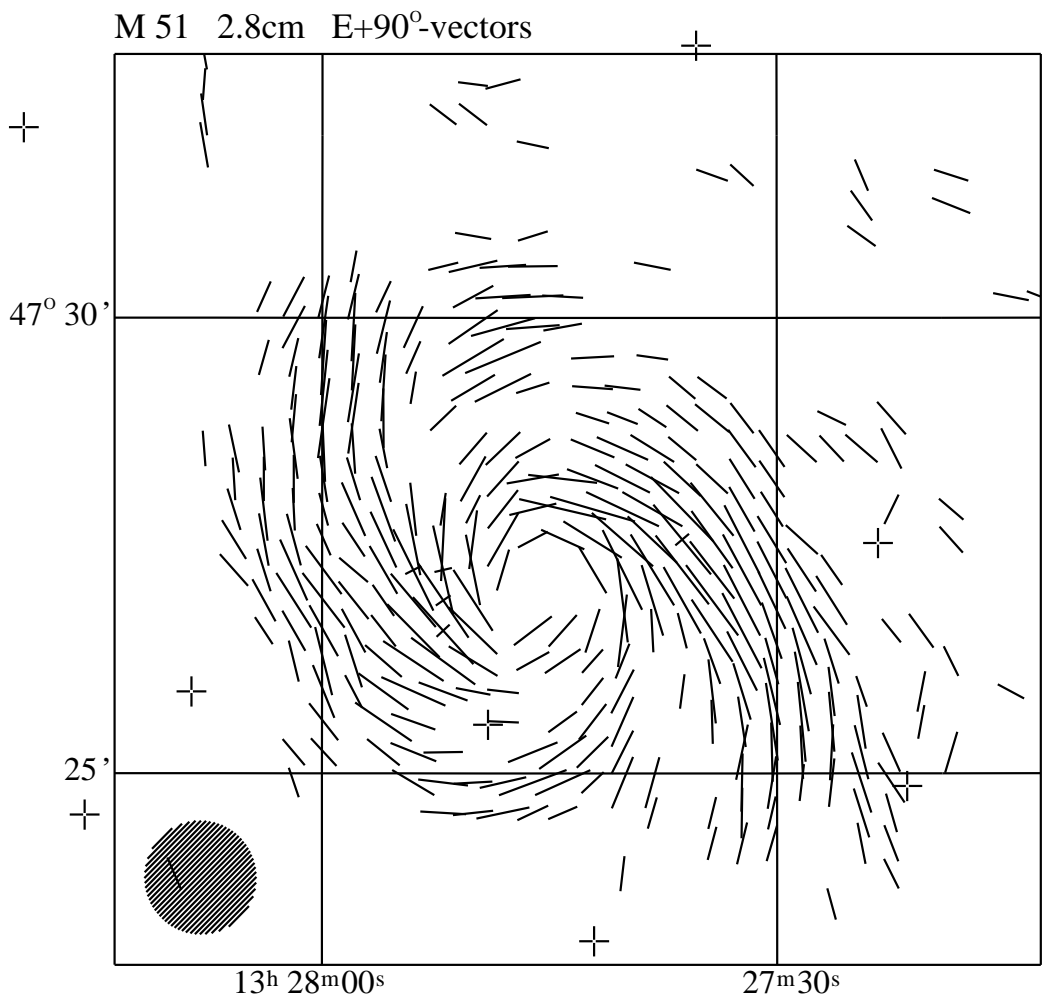
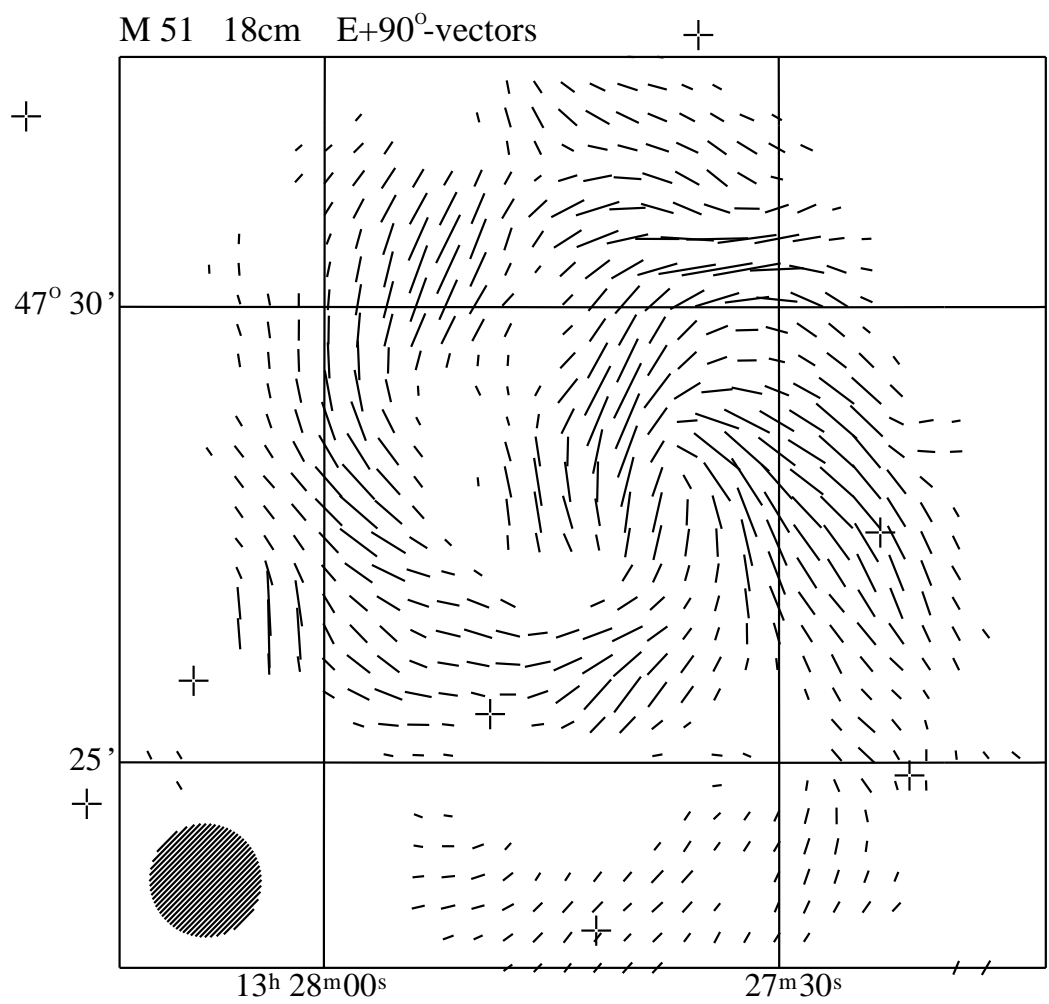
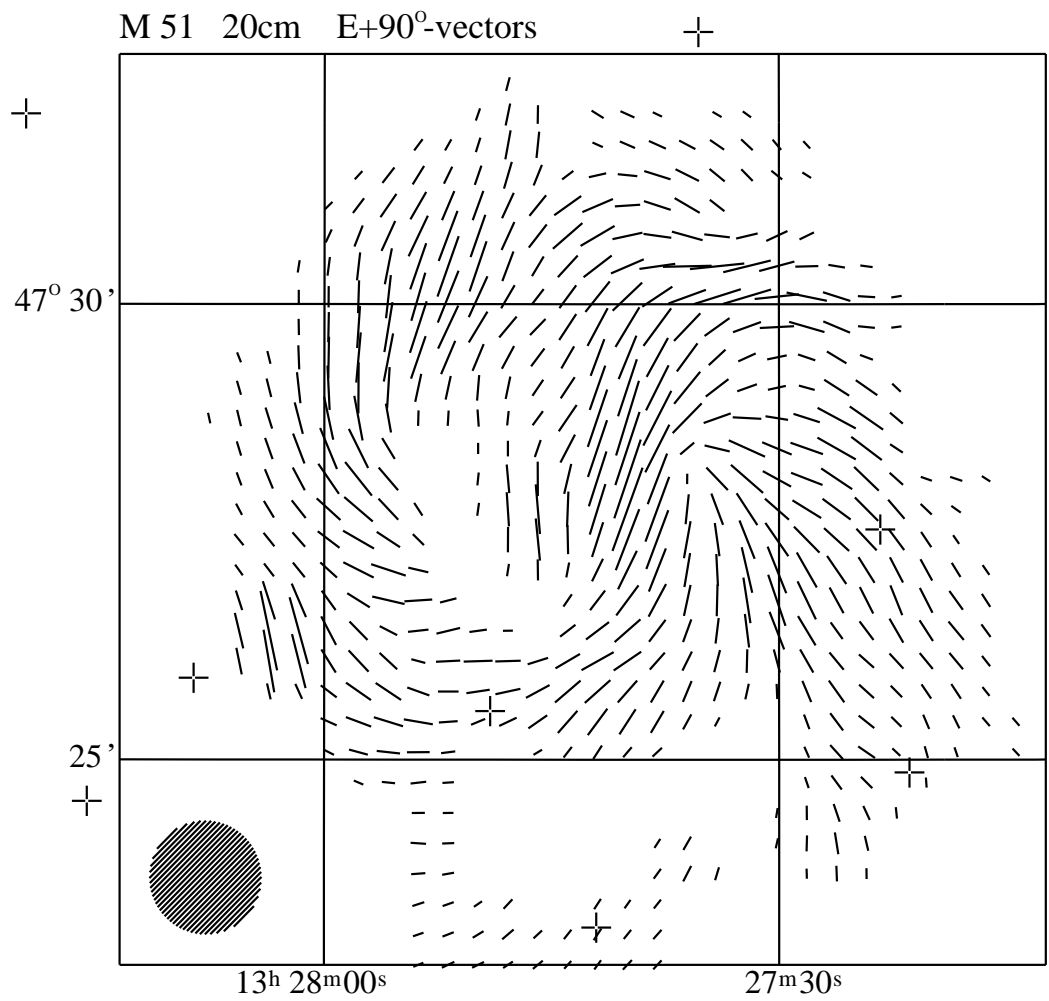
\bye